# A Search for Temporal Changes on Pluto and Charon

J. D. Hofgartner*[1], B. J. Buratti[1], S. L. Devins[1], R. A. Beyer[2,3], P. Schenk[4], S. A. Stern[5], H. A. Weaver[6], C. B. Olkin[5], A. Cheng[6], K. Ennico[3], T. R. Lauer[7], W. B. McKinnon[8], J. Spencer[5], L. A. Young[5], and the New Horizons Science Team

*Corresponding author (Jason.D.Hofgartner@jpl.nasa.gov);

1. Jet Propulsion Laboratory, California Institute of Technology, Pasadena, CA; 2. Sagan Center at the SETI Institute, Mountain View, CA; 3. NASA Ames Research Center, Moffett Field, CA; 4. Lunar and Planetary Institute, Houston, TX; 5. Southwest Research Institute, Boulder, CO; 6. Johns Hopkins University Applied Physics Laboratory, Laurel, MD; 7. National Optical Astronomy Observatory, Tucson, AZ; 8. Washington University in St. Louis, Saint Louis, MO;

Abstract: A search for temporal changes on Pluto and Charon was motivated by (1) the discovery of young surfaces in the Pluto system that imply ongoing or recent geologic activity, (2) the detection of active plumes on Triton during the Voyager 2 flyby, and (3) the abundant and detailed information that observing geologic processes *in action* provides about the processes.  A thorough search for temporal changes using New Horizons images was completed.  Images that covered the same region were blinked and manually inspected for any differences in appearance.  The search included full-disk images such that all illuminated regions of both bodies were investigated and higher resolution images such that parts of the encounter hemispheres were investigated at finer spatial scales.  Changes of appearance between different images were observed but in all cases were attributed to variability of the imaging parameters (especially geometry) or artifacts.  No differences of appearance that are strongly indicative of a temporal change were found on the surface or in the atmosphere of either Pluto or Charon.  Limits on temporal changes as a function of spatial scale and temporal interval during the New Horizons encounter are determined.  The longest time interval constraint is one Pluto/Charon rotation period (~6.4 Earth days).  Contrast reversal and high-phase bright features that change in appearance with solar phase angle are identified.  The change of appearance of these features is most likely due to the change in phase angle rather than a temporal change.  Had active plumes analogous to the plumes discovered on Triton been present on the encounter hemispheres of either Pluto or Charon, they would have been detected.  The absence of active plumes may be due to temporal variability (i.e., plumes do occur but none were active on the encounter hemispheres during the epoch of the New Horizons encounter) or because plumes do not occur.  Several dark streak features that may be deposits from past plumes are identified.

1. Introduction

The New Horizons spacecraft flew through the Pluto-Charon system in July of 2015. Its exploration of the system included imaging with two complementary cameras:

the wide angle Multi-spectral Visible Imaging Camera (MVIC, Reuter et al., 2008) and the narrow angle LOng Range Reconnaissance Imager (LORRI, Cheng et al., 2008). Both imaged Pluto and Charon at orders of magnitude better spatial resolution than any previous observations, resulting in many astounding discoveries (Stern et al., 2015; Moore et al., 2016; Grundy et al., 2016; Gladstone et al., 2016) and enabling a myriad of new scientific investigations. One of the enabled investigations is a search for temporal changes on the surfaces and in the atmospheres from presently active geologic processes at spatial scales that were not previously detectable. In this manuscript, *temporal changes* means any observable changes of appearance over time; this generic terminology is used because the phase space of possible changes of appearance is broad.

Observation of a geologic or atmospheric process *in action* provides abundant and detailed information about the process and this information is sometimes difficult or impossible to derive from observation of the outcome alone. Thus detection of temporal changes on Pluto and Charon is expected to lead to important insights into the geologic or atmospheric processes that are presently operating on these bodies. In fact, characterizing the time variability of Pluto's surface and atmosphere was a secondary objective of the New Horizons mission (Stern 2008). Geologic processes in the solar system, however, are not typically caught in the act and temporal changes are generally more likely to be observed between images with longer time intervals. Due to the flyby architecture of the New Horizons mission, the period of geologically resolved imaging was relatively short, a consideration that decreases the probability that temporal changes were detected.

Nonetheless, there were three reasons for conducting a search for temporal changes:
1. Pluto and Charon have young surfaces and features that imply ongoing or recent geologic activity (Stern et al. 2015),
2. Temporal changes of plumes on Triton were observed in images with similar temporal intervals and spatial resolutions (Soderblom et al., 1990),
3. The significant scientific rewards from detection of temporal changes outweighed the concern of lower return on investment in the event that no temporal changes were detected.

Of the many discoveries from the New Horizons mission, two of the most astounding are the youthfulness of some surfaces and features in the Pluto system and the extent of ongoing and recent geologic activity (Stern et al., 2015). Pluto's Sputnik Planitia[1] has a crater retention age of < 30 million years and resurfacing from convective overturning likely continues to the present (McKinnon et al., 2016; Singer et al., 2017). Directly to its east, eastern Tombaugh Regio also has a young surface that is probably being renewed by actively flowing glaciers (Moore et al., 2016). South of Sputnik Planitia, Wright and Piccard Mons and their surrounding terrain are lightly cratered, possibly because of geologically recent cryovolcanism (Moore et al., 2016). In addition, condensation and sublimation of volatiles is also expected to occur on Pluto and was possibly observed with Earth-based telescopes (e.g., Buie et al., 2010a; Buie et al., 2010b; Buratti et al., 2015).

---

[1] Some feature names in the Pluto-Charon system are presently informal.

On Charon, Organa crater has a concentration of ammonia ice that suggests the ice was deposited < 10 million years ago (Grundy et al., 2016). Formation of Charon's red poles from seasonally cold-trapped volatiles originating from Pluto may be ongoing and thus Charon also is a geologically active world, if only exogenously (Grundy et al., 2016b). The plethora of young surfaces and features in the Pluto system from ongoing and recently active geologic processes imply that observation of a temporal change in New Horizons images was plausible.

Neptune's largest moon, Triton, is regarded as a captured Kuiper belt object and thus is one of the best worlds for comparative planetology for Pluto and Charon. Voyager 2 images from its exploration of the Neptune system in 1989 showed at least four active, geyser-like eruptions of dark material on Triton, two of which had wind-blown clouds that were more than 100 km long (Soderblom et al., 1990). Variability in morphology, brightness, and length of one plume cloud was observed in images with resolutions of ~1 km that were taken less than 45 minutes apart. These observations were used to infer the wind speed and that the eruption was intermittent on 10 minute timescales (Soderblom et al., 1990). New Horizons images of Pluto and Charon at similar spatial resolutions had similar time intervals and thus from the example of Triton, it was plausible that temporal variability was observed.

Temporal changes, including the appearance of dark fans on seasonal ice, that are likely the result of similar geyser-like eruptions have also been observed on Mars (Kieffer et al., 2006; Hansen et al., 2010; Hansen et al., 2013). Mars, Triton, and Pluto are the only solar system bodies that have atmospheres (with surface pressures greater than a microbar) where the most abundant atmospheric component is in vapor pressure equilibrium with surface deposits of its solid phase. The observed activity on both Mars and Triton is likely associated with seasonal sublimation of their surface deposits. Since seasonal sublimation is a process that is also expected to occur on Pluto (e.g., Young 2013) and was possibly observed with Earth-based telescopes, it was predicted that New Horizons would find plume deposits and possibly active plumes (e.g., Buratti et al., 2015). Therefore it was reasonable to search for temporal variability on Pluto that is analogous to that observed on Triton and Mars.

As mentioned previously, the abundant and detailed information that observing a geologic process *in action* provides about that process motivates a search for temporal changes. In the example of Triton's plumes, observation of the active eruptions immediately indicated that the myriad of dark streaks observed on its bright south polar cap were more likely formed by eruptions than saltation, a hypothesis that was given more attention prior to the discovery of the plumes (Sagan and Chyba, 1990). Furthermore, observation of the variability of the plume cloud allowed for detailed measurements such as wind speed and mass flux, which otherwise would not have been uniquely solvable variables. Thus whatever the probability for temporal changes in the brief period of high-resolution New Horizons imaging, the fact that detection of temporal changes was possible and that the scientific rewards for detection of such a change were significant, warranted a search.

2. Data

To search for temporal changes, images with the highest spatial resolution and longest temporal interval are generally the most desirable. Better spatial resolutions permit searches at finer scales and longer time intervals increase the likelihood that changes occurred. In the case of periodic variability, however, it is more optimal to have images at different phases of the period than images separated by many periods but at the same phase. For example, changes driven by diurnal forcing may be more easily detected between two images taken at different times on the same day than two images taken one or more days apart but at the same time of day. For two images with different resolutions, the spatial scale of temporal changes that can be constrained is limited by the lower resolution image.

The New Horizons spacecraft flew through the Pluto system with an approach solar phase angle (Sun-surface-observer angle) asymptote of $\sim 15^0$ and departure of $\sim 165^0$. Since the majority of the visible hemisphere was in nighttime darkness during the departure phase, only the approach and encounter images are useful for searching for temporal changes. Thus the longer the time interval between any two images (from the same camera), the greater the difference of their spatial resolution. Therefore no pair of images optimizes the desire for both high spatial resolution and long temporal interval and every pair of images trades improvement in one parameter at the cost of the other. To be thorough in constraining both the spatial and temporal scales of temporal changes, the highest resolution image should be compared to essentially every other image. Comparison of the highest resolution image to an image taken much earlier provides a long time baseline but only constrains large spatial scales, while comparison to an image taken only slightly earlier significantly improves the constraint of the spatial scale of temporal changes but significantly decreases the time interval.

The problem of selecting the images to compare in the search for temporal changes is further complicated because not all images include the same regions of Pluto/Charon. This is both because of their rotation and because the camera field of view could not always include the whole disk. Thus geographic coverage, in addition to spatial resolution and time interval, must also be considered in the selection of which images to compare. One additional consideration in the selection of images was that New Horizons used two very different imaging cameras, LORRI and MVIC, which have different fields of view and angular resolutions and thus overlap differently in spatial scale and geographic position.

To keep track of the image comparisons, the search for temporal changes on Pluto and Charon was divided into three parts for each body:
1. Full-disk LORRI images such that all regions were investigated (except the part of the southern hemisphere in winter darkness),
2. Higher resolution LORRI images of parts of the encounter hemispheres,
3. MVIC images with information about temporal variability not encompassed by parts 1 and 2.

In this manuscript, *image* is used to refer both to an individual image and a particular sequence of images. Whenever image is used to refer to a sequence, it is a reference to all images in that sequence. We use the New Horizons Sequence ID to refer to specific images (and image sequences). The Sequence ID is the most commonly used name for an image and is included in the header of the image file. Note, however, that for the image comparisons no image mosaics were used, each image in a sequence was compared separately.

The highest resolution LORRI image that included the full visible disk (P_LORRI) was taken as the reference image for Pluto. Table 1 shows the image comparisons for Pluto part 1. All of the other images in the table were acquired prior to P_LORRI and have a lower spatial resolution. These images were selected following two guidelines. One guideline was to avoid a gap of more than a factor of about 2 in the constraint for the spatial scale of temporal changes as a function of the time interval (i.e., the pixel scale of the next image should be less than ~2 times the pixel scale of the previous image). The other guideline was to include images at approximately every 45 degrees of rotation, to ensure that every region was thoroughly investigated for temporal changes (except regions of the southern hemisphere in winter darkness). Table 1 includes the rotational phase relative to P_LORRI (Pluto's rotational period is ~6.4 Earth days) and the pixel scale of each image. A black filled cell indicates that the image at the beginning of the row was compared to the image at the beginning of the column. A white filled cell means that the comparison was not considered necessary based on the two guidelines for selection of image comparisons. Gray filled cells correspond to comparisons that are redundant with their transpose or of an image with itself and thus these cells are not applicable.

Table 1: Image comparisons for part 1 of the search for temporal changes on Pluto. Black filled cells indicate that the images in the row and column header were compared. Gray filled cells at the upper right corner are redundant with their transpose or correspond to comparison of an image with itself. The time of each image is given as a rotational phase relative to P_LORRI (Pluto's rotational period is ~6.4 Earth days) along with the pixel scale of the image in the row headers. The variation of pixel scale within any particular image is less than 10 m. The subsolar and subspacecraft coordinates of P_LORRI are $136^O$ E, $+52^O$ latitude and $159^O$ E, $+40^O$ latitude respectively.

| | PC_MULTI_MAP_A_3 | PC_MULTI_MAP_A_5 | NAV_C4_L1_CRIT_28_01 | PC_MULTIMAP_A_18_L1AH | NAV_C4_L1_CRIT_32_02 | PC_MULTI_MAP_B_3 | PC_MULTI_MAP_B_6 | PC_MULTI_MAP_B_8 | NAV_C4_L1_CRIT_35_02 | NAV_C4_L1_CRIT_36_01 | PC_MULTI_MAP_B_15_01 | P_CLOUD_3_B | PC_MULTI_LONG_1D2A_01 | P_LORRI_TIMERES4 |
|---|---|---|---|---|---|---|---|---|---|---|---|---|---|---|
| **PC_MULTI_MAP_A_3** $\phi\sim-360°-294°$, $S\sim69{,}480$ m | ▨ | ▨ | ▨ | ▨ | ▨ | ▨ | ▨ | ▨ | ▨ | ▨ | ▨ | ▨ | ▨ | ▨ |
| **PC_MULTI_MAP_A_5** $\phi\sim-360°-270°$, $S\sim66{,}960$ m | | ▨ | ▨ | ▨ | ▨ | ▨ | ▨ | ▨ | ▨ | ▨ | ▨ | ▨ | ▨ | ▨ |
| **NAV_C4_L1_CRIT_28_01** $\phi\sim-360°-225°$, $S\sim62{,}170$ m | | | ▨ | ▨ | ▨ | ▨ | ▨ | ▨ | ▨ | ▨ | ▨ | ▨ | ▨ | ▨ |
| **PC_MULTIMAP_A_18_L1AH** $\phi\sim-360°-17°$, $S\sim40{,}360$ m | | | | ▨ | ▨ | ▨ | ▨ | ▨ | ▨ | ▨ | ▨ | ▨ | ▨ | ▨ |
| **NAV_C4_L1_CRIT_32_02** $\phi\sim-360°-1°$, $S\sim38{,}690$ m | | | | | ▨ | ▨ | ▨ | ▨ | ▨ | ▨ | ▨ | ▨ | ▨ | ▨ |
| **PC_MULTI_MAP_B_3** $\phi\sim-320°$, $S\sim34{,}440$ m | ■ | ■ | ■ | ■ | ■ | ▨ | ▨ | ▨ | ▨ | ▨ | ▨ | ▨ | ▨ | ▨ |
| **PC_MULTI_MAP_B_6** $\phi\sim-264°$, $S\sim28{,}590$ m | | ■ | ■ | ■ | ■ | ■ | ▨ | ▨ | ▨ | ▨ | ▨ | ▨ | ▨ | ▨ |
| **PC_MULTI_MAP_B_8** $\phi\sim-227°$, $S\sim24{,}640$ m | | | ■ | ■ | ■ | ■ | ■ | ▨ | ▨ | ▨ | ▨ | ▨ | ▨ | ▨ |
| **NAV_C4_L1_CRIT_35_02** $\phi\sim-179°$, $S\sim19{,}640$ m | | | | ■ | ■ | ■ | ■ | ■ | ▨ | ▨ | ▨ | ▨ | ▨ | ▨ |
| **NAV_C4_L1_CRIT_36_01** $\phi\sim-136°$, $S\sim15{,}130$ m | | | | | ■ | ■ | ■ | ■ | ■ | ▨ | ▨ | ▨ | ▨ | ▨ |
| **PC_MULTI_MAP_B_15_01** $\phi\sim-95°$, $S\sim10{,}860$ m | | | | | | ■ | ■ | ■ | ■ | ■ | ▨ | ▨ | ▨ | ▨ |
| **P_CLOUD_3_B** $\phi\sim-58°$, $S\sim6{,}950$ m | | | | ■ | ■ | ■ | ■ | ■ | ■ | ■ | ■ | ▨ | ▨ | ▨ |
| **PC_MULTI_LONG_1D2A_01** $\phi\sim-27°$, $S\sim3{,}640$ m | | | | ■ | ■ | ■ | ■ | ■ | ■ | ■ | ■ | ■ | ▨ | ▨ |
| **P_LORRI_TIMERES4** $\phi\sim-10°$, $S\sim1{,}890$ m | | | | | | | | | | | | | | ▨ |
| **P_LORRI** $\phi=0°$, $S\sim850$ m | | | | | ■ | ■ | ■ | ■ | | ■ | ■ | ■ | ■ | ■ |

The image comparisons for Pluto parts 2 and 3 are shown in table 2.  All LORRI images acquired after P_LORRI that include Pluto's surface are included in the table using black text.  These images are at a higher resolution than P_LORRI but did not cover the whole disk. P_LORRI_STEREO_MOSAIC covered a significant part of the regions that were covered by the later images and at a higher resolution than P_LORRI and thus was taken as the reference image.  For areas that were not covered by P_LORRI_STEREO_MOSAIC, P_LORRI was the reference image.  These comparisons with P_LORRI for only the areas not covered by P_LORRI_STEREO_MOSAIC are indicated with an orange filled cell in table 2.  P_LORRI and P_LORRI_STEREO_MOSAIC were also compared to each other, which linked parts 1 and 2 so that constraints on the temporal interval extend across the division between the two parts.

MVIC images that were used in Pluto part 3 are included in table 2 using blue text.  These images primarily extended the geographic coverage of the constraints from part 2. The MVIC P_MPAN1 image was the highest resolution image that included the whole disk of Pluto.  Thus it was taken as the reference image and the other MVIC images were compared to it.  The last two MVIC images were compared to the MVIC P_MVIC_LORRI_CA image where they overlapped because that improved the constraint on the spatial scale.  Part 3 was linked to parts 1 and 2 by comparing the MVIC P_MPAN1 image with P_LORRI.  Finally, because the LORRI P_HIPHASE_HIRES covered a part of Pluto that was not illuminated in P_LORRI it was compared to the MVIC P_MVIC_LORRI_CA image.

Table 2: Image comparisons for parts 2 and 3 of the search for temporal changes on Pluto.  Black filled cells indicate that the images in the row and column header were compared.  Orange filled cells indicate that the images in the row and column header were compared only for the regions of Pluto that were not covered by the cell immediately to the right.  Gray filled cells at the upper right corner are redundant with their transpose or correspond to comparison of an image with itself.  Image names in black text are LORRI images and those in blue text are MVIC images.  The time of each image relative to P_LORRI is given in the row headers along with the pixel scale of the image.  The variation of pixel scale within any particular image is less than 10 m.

|  | P_LORRI | P_LORRI_STEREO_MOSAIC | P_LEISA_HIRES | P_MPAN_1 | P_MVIC_LORRI_CA | P_HIPHASE_HIRES | P_LORRI_STEREO_MOSAIC | P_COLOR2 | P_MPAN1 | P_MVIC_LORRI_CA | P_HIPHASE_HIRES | P_CHARONLIGHT |
|---|---|---|---|---|---|---|---|---|---|---|---|---|
|  |  |  |  |  |  |  |  |  |  |  |  |  |

| | P_LORRI | P_LORRI_STEREO_MOSAIC | P_LEISA_HIRES | P_MPAN_1 | P_MVIC_LORRI_CA | P_HIPHASE_HIRES | P_LORRI_STEREO_MOSAIC | P_COLOR2 | P_MPAN1 | P_MVIC_LORRI_CA | P_HIPHASE_HIRES | P_CHARONLIGHT |
|---|---|---|---|---|---|---|---|---|---|---|---|---|
| **P_LORRI** T=0, S~850 m | | | | | | | | | | | | |
| **P_LORRI_STEREO_MOSAIC** T~1.9 hr, S~380 m | ■ | | | | | | | | | | | |
| **P_LEISA_HIRES** T~2.5 hr, S~240 m | ▨ | ■ | | | | | | | | | | |
| **P_MPAN_1** T~3.0 hr, S~120 m | ▨ | ■ | □ | | | | | | | | | |
| **P_MVIC_LORRI_CA** T~3.3 hr, S~80 m | ▨ | ■ | □ | □ | | | | | | | | |
| **P_HIPHASE_HIRES** T~3.7 hr, S~90 m | □ | □ | □ | □ | □ | | | | | | | |
| P_LORRI_STEREO_MOSAIC T~2 hr, S~1440 m | □ | □ | □ | □ | □ | □ | | | | | | |
| P_COLOR2 T~2.8 hr, S~650 m | □ | □ | □ | □ | □ | □ | □ | | | | | |
| P_MPAN1 T~3.0 hr, S~490 m | ■ | □ | □ | □ | □ | □ | ■ | □ | | | | |
| P_MVIC_LORRI_CA T~3.2 hr, S~320 m | □ | □ | □ | □ | □ | ■ | □ | □ | ■ | | | |
| P_HIPHASE_HIRES T~3.7 hr, S~360 m | □ | □ | □ | □ | □ | □ | □ | □ | □ | ▨ | | |
| P_CHARONLIGHT T~3.8 hr, S~450 m | □ | □ | □ | □ | □ | □ | □ | □ | □ | ▨ | ■ | |

Tables 3 and 4 show the image comparisons for Charon in an analogous way to that of tables 1 and 2 for Pluto. C_LORRI was taken as the reference image that links the three parts together. For part 2, C_LEISA_LORRI_1 included all of the areas covered by the two higher resolution images so they were compared to this image, which was compared to C_LORRI. There were only two MVIC images of Charon for part 3 and they were both compared to C_LORRI directly and C_LEISA_LORRI_1 where appropriate.

Table 3: Image comparisons for part 1 of the search for temporal changes on Charon. Black filled cells indicate that the images in the row and column header were compared. Gray filled cells at the upper right corner are redundant with their transpose or correspond to comparison of an image with itself. The time of each image is given as a rotational phase relative to C_LORRI (Charon and Pluto both have rotational periods of ~6.4 Earth days) along with the pixel scale of the image in the row headers. The variation of pixel scale within any particular image is less than 10 m. The subsolar and subspacecraft coordinates of C_LORRI are $315^0$ E, $+52^0$ latitude and $342^0$ E, $+36^0$ latitude respectively.

|  | NAV_C4_L1_CRIT_32_03 | PC_MULTI_MAP_B_3 | PC_MULTI_MAP_B_6 | PC_MULTI_MAP_B_8 | PC_MULTI_MAP_B_11_02 | NAV_C4_L1_CRIT_36_02 | PC_MULTI_MAP_B_15_02 | PC_MULTI_MAP_B_18_02 | PC_MULTI_LONG_1D2A_02 | C_LORRI_TIMERES_1 |
|---|---|---|---|---|---|---|---|---|---|---|
| **NAV_C4_L1_CRIT_32_03** $\phi\sim-361^o$, S~38,740 m | ▒ | ▒ | ▒ | ▒ | ▒ | ▒ | ▒ | ▒ | ▒ | ▒ |
| **PC_MULTI_MAP_B_3** $\phi\sim-320^o$, S~34,470 m | ■ | ▒ | ▒ | ▒ | ▒ | ▒ | ▒ | ▒ | ▒ | ▒ |
| **PC_MULTI_MAP_B_6** $\phi\sim-265^o$, S~28,550 m | ■ | ■ | ▒ | ▒ | ▒ | ▒ | ▒ | ▒ | ▒ | ▒ |
| **PC_MULTI_MAP_B_8** $\phi\sim-227^o$, S~24,570 m | ■ | ■ | ■ | ▒ | ▒ | ▒ | ▒ | ▒ | ▒ | ▒ |
| **PC_MULTI_MAP_B_11_02** $\phi\sim-183^o$, S~19,940 m | ■ | ■ | ■ | ■ | ▒ | ▒ | ▒ | ▒ | ▒ | ▒ |
| **NAV_C4_L1_CRIT_36_02** $\phi\sim-136^o$, S~15,090 m | ■ | ■ | ■ | ■ | ■ | ▒ | ▒ | ▒ | ▒ | ▒ |
| **PC_MULTI_MAP_B_15_02** $\phi\sim-96^o$, S~10,890 m | ■ | ■ | ■ | ■ | ■ | ■ | ▒ | ▒ | ▒ | ▒ |
| **PC_MULTI_MAP_B_18_02** $\phi\sim-61^o$, S~7,250 m | ■ | ■ | ■ | ■ | ■ | ■ | ■ | ▒ | ▒ | ▒ |
| **PC_MULTI_LONG_1D2A_02** $\phi\sim-27^o$, S~3,700 m | ■ | ■ | ■ | ■ | ■ | ■ | ■ | ■ | ▒ | ▒ |
| **C_LORRI_TIMERES_1** $\phi\sim-6^o$, S~1,540 m |  |  |  |  |  |  |  |  |  | ▒ |
| **C_LORRI** $\phi=0^o$, S~860 m | ■ | ■ | ■ |  | ■ | ■ | ■ | ■ | ■ |  |

Table 4: Image comparisons for parts 2 and 3 of the search for temporal changes on Charon. Black filled cells indicate that the images in the row and column header were compared. Orange filled cells indicate that the images in the row and column header were compared only for the regions of Charon that were not covered by the cell immediately to the right. Gray filled cells at the upper right corner are redundant with their transpose or correspond to comparison of an image with itself. Image names in black text are LORRI images and those in blue text are MVIC images. The time of each image relative to C_LORRI is given in the row headers along with the pixel scale of the image.  The variation of pixel scale within any particular image is less than 10 m.

|  | C_LORRI | C_LEISA_LORRI_1 | C_LEISA_HIRES | C_MVIC_LORRI_CA | C_COLOR2 | C_MVIC_LORRI_CA |
|---|---|---|---|---|---|---|
| **C_LORRI** T=0, S~860 m |  |  |  |  |  |  |
| **C_LEISA_LORRI_1** T~0.7 hr, S~690 m | ■ |  |  |  |  |  |
| **C_LEISA_HIRES** T~1.9 hr, S~410 m |  | ■ |  |  |  |  |
| **C_MVIC_LORRI_CA** T~3.2 hr, S~160 m |  | ■ |  |  |  |  |
| **C_COLOR2** T~2.1 hr, S~1430 m | ■ |  |  |  |  |  |
| **C_MVIC_LORRI_CA** T~3.2 hr, S~620 m | 🟧 | ■ |  |  |  |  |

The comparisons in tables 1-4 show that the search for temporal changes was indeed thorough but that not every possible part of the spatial scale, temporal interval, geographic coverage phase space that could have been investigated was investigated.  Additional comparisons would only modestly improve the constraints on temporal changes and were not pursued exhaustively as the effort would exceed the return.  Conversely, a few of the comparisons in tables 1-4 were not necessary as their constraint is fully encompassed by other comparisons.  These comparisons were extra checks for evidence of temporal changes.

3. Methods

To search for temporal changes, images that covered the same region were blinked and manually inspected for any differences of appearance of the surface or atmosphere (e.g., appearance of a cloud).  To facilitate comparison of images, the images were adjusted to either have the same camera geometry or map projection using the United States Geological Survey's (USGS) Integrated Software for Imagers and Spectrometers (ISIS) cam2cam and cam2map functions with cubic convolution interpolation.  The blink comparisons were at the pixel scale of the lower resolution image.  To reduce the possibility of accidentally missing a difference of appearance, two researchers among us (JDH and SLD) checked each image pair.  The blink comparisons were systematically checked for all of the comparisons shown in tables 1-4.

An automated change detection algorithm is conceivable but manual inspection of the image pairs was favored for several reasons. As a result of the flyby architecture of the New Horizons mission, the imaging geometry varied drastically between some images. Thus changes of appearance from shadowing, foreshortening, parallax, etc. were common, especially for the rugged regions of Pluto and Charon. These changes of appearance could easily be eliminated from consideration as a temporal change by a manual inspector but an algorithm would flag them all. Secondly, slight errors in the co-registration of the images would cause an algorithm to flag many false-positive changes of appearance whereas a manual inspector can also easily distinguish these offsets and ignore them while comparing the images. Thirdly, since many different active processes are conceivable (especially in the first exploration of a new class of planetary bodies) and any temporal change is of interest, the phase space of conceivable changes in appearance is broad and it would be difficult to implement an algorithm that would be as sensitive to the whole phase space as a manual inspector. For these three reasons, the images were manually searched for temporal changes, despite the extra labor involved.

4. Results

Changes in appearance between different images were frequently observed but in all cases could be attributed to variability of the imaging parameters (especially geometry) or artifacts. No difference of appearance that is strongly indicative of a temporal change was found on the surface or in the atmosphere of either Pluto or Charon. The constraints on temporal changes on Pluto and Charon during the New Horizons encounter from this search cannot be succinctly summarized as they depend on the geographic position, spatial scale, and temporal interval. Nevertheless, some generalized constraints can be derived from each part of the search.

On Pluto's encounter hemisphere, there were no temporal changes > 1 km in scale over ~3 hour period (part 3), > 2 km over ~7 hours (parts 1 and 3), and > 40 km over a full rotation period (~6.4 Earth days; part 1). On Pluto's non-encounter hemisphere, there were no temporal changes > 63 km in scale over a full rotation period (part 1). For ~5% of Pluto's surface, there were no temporal changes > 400 m in scale over ~1 hour period (part 2). On Charon's encounter hemisphere, there were no temporal changes > 1.5 km in scale over ~5 hour period (parts 1 and 3) and > 40 km over a full rotation period (part 1). On Charon's non-encounter hemisphere, there were no temporal changes > 35 km in scale over ~3/8 of a rotation period (~2.4 Earth days; part 1). For ~23%, of Charon's surface, there were no temporal changes > 700 m, over ~2.5 hour period (part 3). These constraints are shown in Table 5.

Table 5: Some generalized constraints on temporal changes on Pluto and Charon during the New Horizons encounter. There were no temporal changes in the parameter spaces indicated in the rows. This table does not include all of the constraints from this search.   The constraints from this search depend on the geographic position, spatial scale, and temporal interval and cannot all be succinctly summarized.

| Region | Spatial Scale | Temporal Interval |
|---|---|---|
| Pluto encounter hemisphere | > 1 km | < 3 hr |
| Pluto encounter hemisphere | > 2 km | < 7 hr |
| Pluto encounter hemisphere | > 40 km | < 1 Pluto day (~6.4 Earth days) |
| ~5% of Pluto's surface | > 400 m | < 1 hr |
| Pluto non-encounter hemisphere | > 63 km | < 1 Pluto day |
| Charon encounter hemisphere | > 1.5 km | < 5 hr |
| Charon encounter hemisphere | > 40 km | < 1 Charon day |
| ~23% of Charon's surface | > 700 m | < 2.5 hr |
| Charon non-encounter hemisphere | > 35 km | < 3/8 Charon day (~2.4 Earth days) |

In total, 165 individual images were considered in this search for temporal changes.  Two or more changes in appearance were observed in most image comparisons.  Thus several hundred changes in appearance were identified and it would be too cumbersome to list them in this manuscript.  Three of the most interesting examples of changes of appearance between images, which are attributed to changes of imaging parameters or artifacts rather than temporal changes on Pluto or Charon, are shown below.

4.1 Contrast Reversal Features on Pluto

Contrast reversal features are an interesting example of a change of appearance between different images that is not due to a temporal change but rather to a change of the imaging geometry.  Figure 1 shows three images of the same region of Sputnik Planitia on Pluto.  In the left panel red arrows indicate three features that are darker than their bright surroundings in an image taken at a solar phase angle of ~$25^O$.  In the middle image, taken at a phase angle of ~$47^O$, one of the features is barely apparent and the other two are not observed.  In the right image, taken at a phase angle of ~$151^O$, all three features are brighter than their surroundings.  Thus these features have reversed in contrast relative to their surroundings as the phase angle increased.  When the brightness of the middle image is very strongly stretched, all three features are observed.  Not included in the figure are images at phase angles of ~$19^O$, $70^O$, and $161^O$ that exhibit the same respective behavior as the images in figure 1.  At a solar phase angle of ~$37^O$ the presence of the features is variable in the different MVIC color filters.  LORRI images at about four times better resolution (~100 m pixel scale) than the MVIC images at $47^O$ and $70^O$ also covered this region and show similar behavior to their corresponding MVIC images.

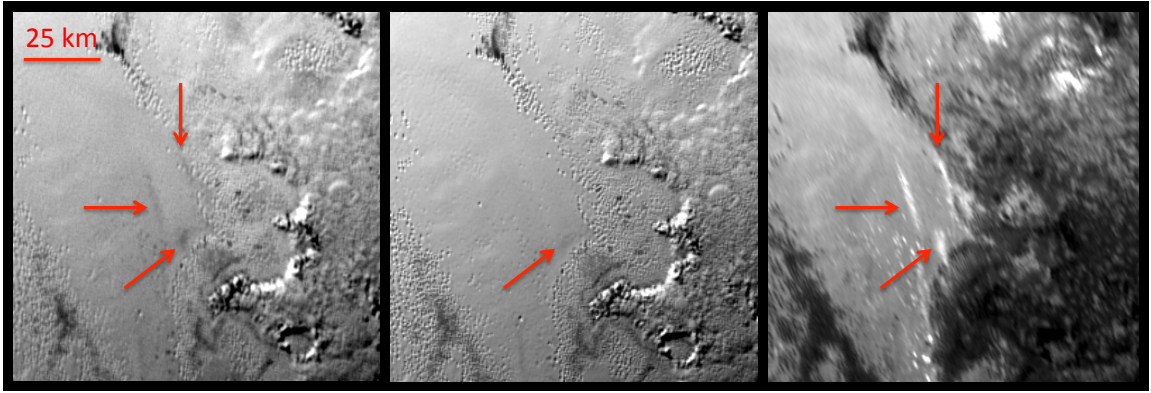

Figure 1: Contrast Reversal Features. The areas indicated with red arrows change in appearance between these three images of part of Pluto's Sputnik Planitia. They are contrast reversal features that are dark relative to their surrounding terrain at low phase (25$^0$, left panel), not/barely apparent at intermediate phase (47$^0$, middle panel), and brighter than their surroundings at high phase (151$^0$, right panel). The change of appearance is due to the change of phase angle rather than a temporal change on Pluto. The darkening of regions at the right and bottom left of the right panel is also due to the high phase angle, most likely shadowing, and the bright features at the top right are additional examples of contrast reversal features.

These features were stationary over the ~4 hour period when images that could resolve them were acquired. Thus, while the features are difficult to see in the images at intermediate phase, and reverse in contrast relative to their surroundings in the images at high phase, they were present when all of the images were acquired. Therefore the change of appearance of the features is more likely due to the change of phase angle than a temporal change.

Additional contrast reversal features are observed in figure 1 and elsewhere on Pluto. Hypotheses for the contrast reversal features include deposits from former plumes, haze clouds, and anomalously smooth surfaces. Further investigation into the nature of these intriguing features and the details of the cause of their change in appearance with phase angle may confirm or refute these hypotheses. For the purposes of this project to search for temporal changes, however, it is concluded that the change of appearance is not due to a temporal change.

4. 2 High-phase Bright Features on Pluto

Another example of changes of appearance between images that are attributed to changes in the phase angle rather than temporal changes are high-phase bright features. These features are similar to the contrast reversal features in that they appear brighter than their surroundings at high phase angles and are not observed at intermediate phase angles. However, they are also not observed at low phase angles whereas the contrast reversal features are observed to be darker than their surroundings.

Figure 2 shows three images of the same region of Pluto's Sputnik Planitia that includes several high-phase bright features. These features are common on Pluto in areas with volatile deposits that appear bright and smooth and do not exhibit pits. But not all areas that match this description at low phase angle have high-phase bright features (see figure 2) and there are a few areas with pits that do brighten in high phase angle images. The abundance of these features and their detection in only high phase angle images indicates that they are more likely due to changes of the imaging geometry than temporal changes on Pluto.

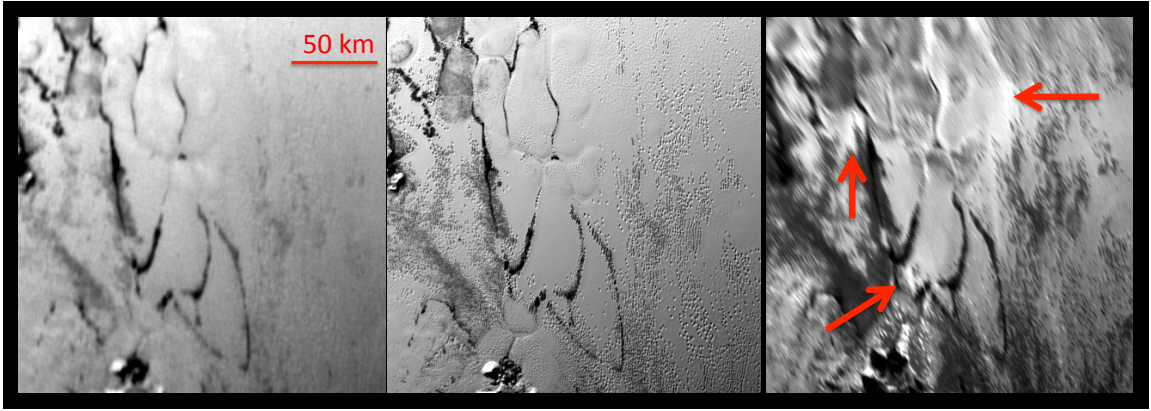

Figure 2: High-phase Bright Features. The red arrows indicate some areas of Pluto's Sputnik Planitia that change in appearance between the right panel and the other two panels. Bright features are observed at high phase angle ($151^0$, right panel) but not in any lower phase angle images ($19^0$, left panel; $69^0$, middle panel; $25^0$, $38^0$, and $47^0$, not shown). Other areas in these images also exhibit the same behavior. The change of appearance is due to the change of phase angle rather than a temporal change on Pluto. The darkening of many areas in the right panel is also due to the high phase angle, most likely shadowing.

The angle between specularly reflected rays from a smooth surface and the vector from the surface to the camera (scattering angle) varies between ~$8^0$-$15^0$ for the high-phase bright features in the right panel of figure 2. Thus these features may be reflections from smooth surface facets with slopes of ~$8^0$-$15^0$.

4.3 Spatially-coincident Cosmic Rays

Aside from changes of the imaging geometry, image artifacts also resulted in changes of appearance between images of the same region. One of the more dramatic examples of artifacts that caused a change of appearance is the spatially-coincident cosmic rays shown in figure 3. The top middle and top right images of figure 3 show bright features at nearly the same location on a slope that are indicated by red arrows. The bright features, however, are not observed in an image acquired ~14 minutes prior (top left) nor in an image acquired at nearly the same time as the top middle and right images (bottom left). Both of these images, however, do have sufficient spatial resolution to detect the bright feature in the top right image had it been present. Artifacts that cause a change of appearance between images are not uncommon, especially cosmic ray artifacts, but it is rare that they change the appearance of two images at nearly the same location and thus these features warranted further investigation.

The bottom middle and bottom right panels of figure 3 show the counts value after flat field calibration for every pixel of the images that are shown directly above them (only part of the images are shown). That is, they are one-dimensional plots of the brightness of each pixel in the image in camera units. Red arrows in the plot panels indicate the pixels that are indicated by red arrows in the image panels directly above them. The bright features clearly standout as spikes in the plots. Many temporal changes could manifest as a bright feature but are not necessarily a priori expected to be extraordinarily brighter than all other nearby features. Thus, that these features correspond to spikes in the brightness plots, suggests that they are artifacts rather than a temporal change.

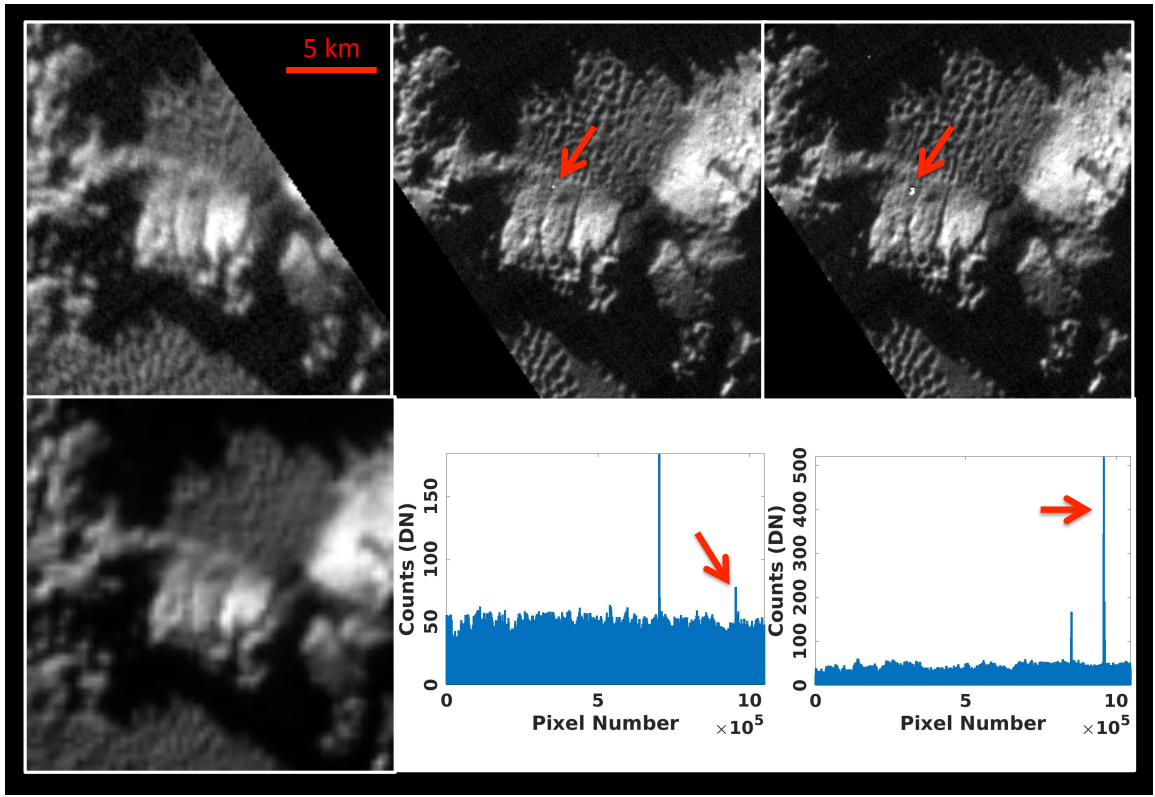

Figure 3: Spatially-coincident Cosmic Rays. There are bright features in the top middle and top right images, indicated by red arrows that are not observed in the two left images. This change of appearance is attributed to cosmic rays artifacts in the top middle and right images that occurred at nearly the same geographic location on Pluto. The bottom middle and bottom right panels are one-dimensional plots of the counts value for every pixel of the images that are shown directly above

them (only part of the images are shown). The bright features in the images correspond to spikes in the plots, indicated by red arrows in both the images and plots. This suggests that the bright features are artifacts, most likely cosmic rays, and not due to a temporal change on Pluto. The four images shown are from LOR_0299178925, LOR_0299179751, LOR_0299179754, and MP2_0299179552.

Upon inspection of brightness plots for many more LORRI images, it was found that most have 1 or 2 spikes. Such spikes look like cosmic ray artifacts, do not appear to cast shadows, are generally not saturated pixels, and are randomly distributed on the focal plane array. Thus the bright features in figure 3 are attributed to cosmic rays that coincidentally occurred at approximately the same geographic location on Pluto. Note that the cosmic rays occurred at approximately the same geographic location on Pluto but not at the same location of the array on the focal plane, they occurred in different pixels in the two images. They are separated by ~500 m (pixel scale of both images is ~76 m) between the two images, further supporting the conclusion that they are cosmic ray artifacts. The two spikes in the bottom middle and bottom right panels of figure 3 that are not indicated by red arrows are similarly attributed to cosmic ray artifacts.

5. Discussion

Active plumes were observed during the Voyager 2 encounter with Triton, the most similar body to Pluto that has been explored by a spacecraft, but the results of this work indicate that no similar active plumes were observed during the New Horizons encounter with Pluto. New Horizons images of Pluto included images that had similar observational parameters (such as spatial resolutions, temporal intervals, solar phase angles, etc.) to the Voyager 2 images that detected Triton's plumes and thus the difference in observation of active plumes is unlikely to be an observational effect. Is the difference related to the timing (ex. diurnal or seasonal) of the observations or are the plumes phenomena that occur on Triton but not Pluto?

At least four plumes were active during the Voyager 2 encounter with Triton (Soderblom et al., 1990), and more than 100 dark surface streaks that are expected to be the result of previous plume activity were mapped (Hansen et al., 1990). All of these features were superposed on Triton's south polar cap. The cap covered the southern hemisphere up to its margin, which ranged with longitude from $30^0$ S to $10^0$ N. The active plumes were at latitudes from $49^0$ S to $60^0$ S while the dark surface streaks were observed at latitudes from $5^0$ S to $70^0$ S and most abundant from $15^0$ S to $45^0$ S (Hansen et al., 1990). The features were distributed in longitude and ranged in length from ~4 km to > 100 km. At the time of the Voyager 2 encounter, the subsolar latitude on Triton was $45^0$ S and moving southward to solstice at $52^0$ S. The latitudes of the plumes and streaks relative to the subsolar latitude is suggestive of a solar driven process and a model where solar heating drives geyser-like eruptions has been proposed (e.g., Soderblom et al., 1990; Kirk et al., 1995).

The results of this work provide constraints on temporal changes for the entire encounter hemisphere of Pluto.  Since at any point in time, different longitudes correspond to different times of day, temporal variability is effectively constrained for a range of times of day.  Thus it is unlikely that diurnal variability of plume activity is the reason that active plumes were observed on Triton but not Pluto.

The subsolar latitude on Pluto during the New Horizons encounter was +52$^O$ latitude and moving northward to solstice at +60$^O$ latitude.  This is slightly north of the volatile filled basin Sputnik Planitia, which has a maximum latitude of +50$^O$ and near the southern extent of the north polar cap Lowell Regio.  From direct analogy to Triton, active plumes would be expected on Lowell Regio and dark surface streaks would be expected on Sputnik Planitia.

Active plumes were not observed in either region or anywhere else on Pluto.  Features that are darker than their surroundings and streak-like in shape, however, abound on the surface.  Most do not resemble the dark surface streaks on Triton that were interpreted to be the result of previous plume activity because they appear to be controlled by topography and/or tectonics (ex. troughs of Sputnik Planitia cells and faults), are adjacent to hills and are more likely to be caused by interactions between the hills and winds than an eruptive process (Stern et al., 2015), are not as dark, or have branching or sinuosity and thus are less likely to be wind blown.

A few of the features that most resemble in appearance the dark surface streaks on Triton are shown in figure 4.  The features in Sputnik Planitia indicated by red arrows in panels A-D may be smaller examples of the streaks that are adjacent to hills where the hill is not resolved in these cases.  All of the Sputnik Planitia streaks that are adjacent to a resolvable hill, however, originate at a cell boundary, whereas the three features that are indicated by red arrows in panels A-C originate in the interior of a cell.  The contrast reversal features indicated by red arrows in figure 1 are also dark streak features (at low solar phase angles) on Sputnik Planitia that are not adjacent to a hill or a cell boundary.  The length of all of these features is similar to the shortest of the dark surface streaks that were mapped on Triton.  Panel E of figure 4 shows three dark streak features to the east of Sputnik Planitia.  The long axes of these features are all approximately aligned with nearby pit chains suggesting that these features may have some tectonic control and thus may not be wind-blown deposits from former plumes.  Several dark streak features are observed in and near Burney crater (shown in panel F, only a few of the many dark streak features are indicated with green arrows).

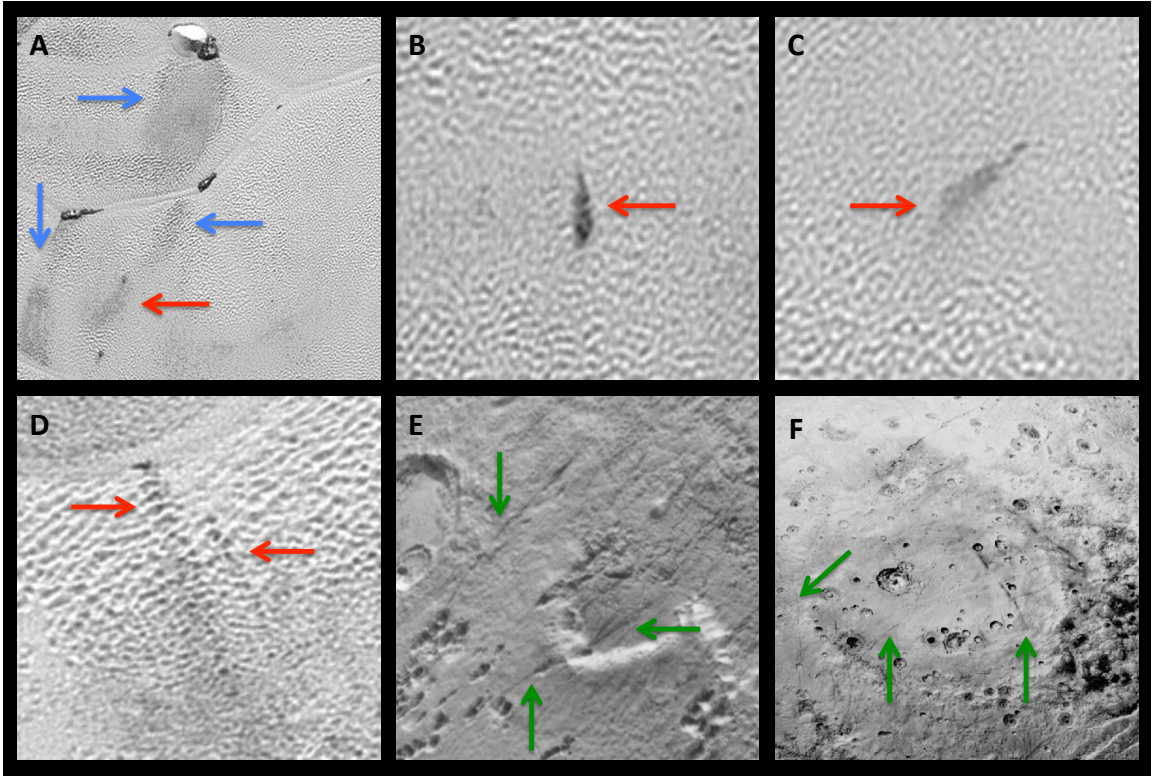

Figure 4: A few examples of dark surface streaks on Pluto. Blue arrows in panel A indicate streaks in Sputnik Planitia that are adjacent to hills and may be the result of interactions between the hills and winds (Stern et al., 2015). Red arrows in panels A-D indicate a few streaks in Sputnik Planitia that may be smaller examples of the streaks indicated by blue arrows in panel A or that may be the result of previous plumes. Green arrows in panel E indicate three dark streaks to the east of Sputnik Planitia that may be tectonically controlled or the result of previous plumes. Green arrows in panel F indicate a few of the many dark streaks in and near Burney crater. Panel A is ~70 km by 70 km, panels B and C are ~16 km by 16 km, panel D is ~18 km by 18 km, panel is E is ~150 km by 150 km, and panel F is ~450 km wide by 350 km high.

In the case that some of the dark streak features in figures 1 and 4 are the result of previous plumes, seasonal variability of the plumes may explain the lack of active plumes observed during the New Horizons encounter. It is possible that the timing of the encounter was during a period when seasonal Sputnik Planitia plumes have finished and Lowell Regio plumes have not yet started. Further analysis of Pluto's dark surface streaks may confirm or refute this possibility.

Diurnal variability and differences in observational parameters are unlikely to be the reason that active plumes were observed on Triton but not Pluto. Seasonal variability is a possible explanation for this difference but an additional possibility to consider is that the process that generates the plumes does not occur on Pluto.

Sublimation of nitrogen ice due to solar heating in a sealed subsurface greenhouse environment (solid-state greenhouse effect) followed by explosive eruption of the pressurized nitrogen gas is the leading hypothesis for the mechanism that generates Triton's plumes (Soderblom et al., 1990; Kirk et al., 1995). In that model, Triton's south polar cap is primarily high albedo nitrogen ice but has dark material that is near enough to the surface to absorb incident solar energy and heat the subsurface nitrogen ice. This heating increases the vapor pressure of nitrogen but the nitrogen gas is sealed within the ice cap. The gas pressure builds until it explosively erupts to the surface bringing dark particles with it that constitute the observable plume. Similar processes with carbon dioxide substituting for nitrogen are likely responsible for abundant temporal variability observed in Mars' polar regions (Kieffer et al., 2006; Hansen et al., 2010; Hansen et al., 2013).

Pluto's Sputnik Planitia is also a nitrogen-ice cap that solar photons penetrate relatively deeply (Grundy et al., 2016; Protopapa et al., 2017). The surface has an extremely high albedo (Buratti et al., 2017) but dark material is observed in cell troughs and pit floors (Moore et al., 2016). Insufficient mixing of the dark material and nitrogen ice, however, is one speculative difference that may result in insufficient subsurface heating and sublimation to generate explosive eruptions. Other hypothetical differences, such as more internal fracturing in Sputnik Planitia, so that nitrogen gas vents to the surface as it sublimates rather than accumulating and increasing in pressure, may explain the absence of plumes. The spectrum of Lowell Regio shows little absorption from nitrogen, due to a depletion of nitrogen or shorter optical path lengths, either of which could prevent solar-driven nitrogen eruptions (Grundy et al., 2016; Protopapa et al., 2017). Thus differences of the volatile caps on Pluto and Triton may prevent the active plumes that were observed on Triton from occurring on Pluto.

An alternative model for Triton's plumes involves nitrogen ice melting at depth due to internal heat flow (Brown and Kirk, 1994). In this case Triton's greater internal heat flow, augmented with respect to that of Pluto due to Triton's larger size and modest obliquity tides (Nimmo and Spencer, 2015), may be responsible for the presence of active plumes on Triton and their apparent absence on Pluto.

Clyde Tombaugh discovered Pluto by blinking between different photographic plates of the same regions of the sky and inspecting them for any differences that would correspond to the orbital motion of a trans-Neptunian object (TNO). The reward of his systematic and meticulous search was the discovery of Pluto in 1930. We followed in Tombaugh's footsteps by blinking between different images of the same regions of Pluto and Charon, also in the hope of informative discoveries in the distant solar system.

No changes in appearance were found in New Horizons images that are strongly indicative of a temporal change. This result is likely due to the relatively brief period of geologically resolved imaging, not an absence of ongoing geologic processes in the Pluto-Charon system. The youthfulness of several surface units (e.g., Moore et al., 2016), the observation of temporal changes over longer time intervals with Earth-based telescopes (e.g., Buratti et al., 2015), and predictions for significant changes in Pluto's atmospheric pressure (e.g., Stern et al., 2017) all indicate that geologic processes are presently active in the Pluto-Charon system. Thus it is very probable that ongoing geologic activity will result in changes that are detected with future observations. In particular an orbiter mission that probes temporal variability on multiple timescales at multiple spatial scales is likely to observe changes.

6. Conclusions

A search for temporal changes on Pluto and Charon using New Horizons images was completed. No changes in appearance that are strongly indicative of a temporal change were found on the surface or in the atmosphere of either Pluto or Charon. All illuminated regions of both bodies were investigated.

On Pluto's encounter hemisphere, there were no temporal changes > 1 km in scale over ~3 hour period, > 2 km over ~7 hours, and > 40 km over one rotation period (~6.4 Earth days). On Pluto's non-encounter hemisphere, there were no temporal changes > 63 km in scale over one rotation period. For ~5% of Pluto's surface, there were no temporal changes > 400 m in scale over ~1 hour period. On Charon's encounter hemisphere, there were no temporal changes > 1.5 km in scale over ~5 hour period and > 40 km over one rotation period. On Charon's non-encounter hemisphere, there were no temporal changes > 35 km in scale over ~3/8 of a rotation period. For ~23%, of Charon's surface, there were no temporal changes > 700 m, over ~2.5 hour period.

Changes in appearance between different images were observed but in all cases were attributed to variability of the imaging parameters (especially geometry) or artifacts. Contrast reversal and high-phase bright features that change in appearance with solar phase angle were identified. The change of appearance of these features is most likely due to the change in phase angle rather than a temporal change.

Had active plumes analogous to the plumes discovered on Triton been present on the encounter hemispheres of either Pluto or Charon, they would have been detected. Several dark streak features that may be deposits from past plumes were identified. The absence of active plumes may be due to seasonal variability or because the process that generates the plumes does not occur on Pluto.

Based on the youthfulness of several surface units in the Pluto system; it is probable that ongoing geologic activity will result in changes that are detected with future observations. An orbiter mission that probes temporal variability on multiple timescales at multiple spatial scales is likely to observe changes.

Acknowledgements: Jason Hofgartner gratefully acknowledges the entire New Horizons team for the incredible exploration of the Pluto system that enabled this research.  Jason Hofgartner was supported by an appointment to the NASA Postdoctoral Program at the NASA Jet Propulsion Laboratory, California Institute of Technology administered by Universities Space Research Association through a contract with NASA.  This research has made use of the USGS Integrated Software for Imagers and Spectrometers (ISIS).  We thank two anonymous reviewers for their service and helpful comments.